\documentclass{ws-ijmpc}

\begin{document}

\markboth{B.~Mitrovi\' c and M.~A.~Przedborski}
{Monte Carlo Study of the XY-model on Sierpi\' nski Carpet}

%\catchline{}{}{}{}{}
%%%%%%%%%%%%%%%%%%%%%%%%%%%%%%%%%%%%%%%%%%%%%%%%%%%%%%%%%%%%%%%%%%%%

\title{MONTE CARLO STUDY OF THE XY-MODEL ON SIERPI\' NSKI CARPET}

\author{BO\v ZIDAR MITROVI\'C}

\address{Department of Physics, Brock University, 500 Glenridge Ave.\\
St.~Catharines, Ontario L2S 3A1, Canada\\
bmitrovic@brocku.ca}

\author{MICHELLE A. PRZEDBORSKI}

\address{Department of Physics, Brock University, 500 Glenridge Ave.\\
St.~Catharines, Ontario L2S 3A1, Canada\\
mp06lj@badger.ac.brocku.ca}

\maketitle

\begin{history}
\received{Day Month Year}
\revised{Day Month Year}
\end{history}

\begin{abstract}
We have performed a Monte Carlo study of the classical XY-model on a Sierpi\' nski carpet,  
which is a planar fractal structure with infinite order of ramification and fractal
dimension 1.8928. We employed the Wolff cluster algorithm in our simulations and our results, in 
particular those for the susceptibility and the helicity modulus, indicate the absence of finite-temperature 
Berezinskii-Kosterlitz-Thouless (BKT) transition in this system. 

\keywords{Fractals; XY-model; Monte Carlo}
\end{abstract}

\ccode{PACS Nos.: 64.60.al, 64.60.Bd, 64.60.De}

\section{Introduction}

In a translationally invariant system the critical behavior is determined by its universality class 
which depends on the symmetry of the Hamiltonian, the spatial dimension of the system, and the 
range of forces between the particles. Systems with fractal structure lack translational invariance but 
are scale invariant and have additional topological features, such as the fractal dimension, the order of 
ramification, the connectivity, and the lacunarity, which could influence the phase transition in 
these systems (see References \refcite{gab80}--\refcite{mh}). Much is known about the phase 
transitions for the discrete-symmetry Ising model on various fractal structures  
(Refs. \refcite{gab80,gab83,gasb84,gab84,mh}) but very little is known about the continuous-symmetry models,
such as the classical XY-model, on fractals\cite{vkb}. 
In a recent work we used the Metropolis Monte Carlo (MC) method to examine the classical XY-model on two  
fractal structures with finite orders of ramification, Sierpi\' nski gasket\cite{mb10} and
Sierpi\' nski pyramid\cite{pm11}, and found no phase transition at any finite temperature. 
These results were analogous to what was obtained previously\cite{gab80,gab83,gasb84,gab84} 
for the Ising model on finitely ramified fractals using the renormalization group method.
The order of ramification $R$ at a point $P$ of a structure is defined as the number of bonds which
must be cut in order to isolate an {\em arbitrarily large} bounded set of points connected to $P$ 
(e.g.~the regular periodic lattices have $R=\infty$). In systems with finite $R$ the thermal fluctuations 
at any finite temperature destroy long-range order (no finite-temperature continuous phase transition), 
as well as quasi-long-range order (no finite-temperature Berezinskii-Kosterlitz-Thouless (BKT) transition).
These conclusions hold regardless of the symmetry of the microscopic Hamiltonian 
(i.e. whether it has a discrete $Z_{2}$ symmetry (Ising) or a continuous $O(2)$ symmetry (XY)).  

For discrete-symmetry spin models on infinitely ramified Sierpi\' nski carpets, Gefen, Aharony,  
and Mandelbrot\cite{gab84} found phase transitions at finite temperature (see also the work of
Wu and Hu\cite{wh87} where some of the recursion relations in Ref.~\refcite{gab84} have been corrected). 
Moreover, using a correspondence between a resistor network connecting the sites of a given
lattice and the low-temperature properties of $n$-component spin models, with $n\geq$ 2, on the same  
lattice\cite{stinc79} they argued that there is no long-range order at any finite temperature if the 
fractal dimension $D<$ 2, even when $R=\infty$. They conjectured that the reason for the absence of the 
long-range order is that for $O(n)$ spin models the lower critical Euclidean dimension is $d=$ 2, and 
for a fractal in $d$ dimensions one has $D\leq d$. 

\begin{figure}[ph]
\centerline{\psfig{file=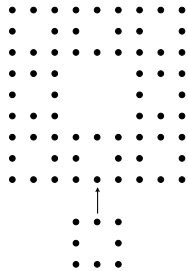,width=4.7cm}}
\vspace*{8pt}
\caption{The first step in creating Sierpi\' nski carpets with $b$=3 and $l$=1.}
\label{fig1}
\end{figure}

Here we present a Monte Carlo study of the classical XY-model on a two-dimensional Sierpi\' nski carpet 
depicted in Figure~\ref{fig1}, described by the Hamiltonian
\begin{equation}
\label{eq:ham}
 H=-J\sum_{\langle i,j\rangle}\cos(\theta_{i}-\theta_{j})\>,
\end{equation}
where 0 $\leq \theta_{i}<$ 2$\pi$ is the angle variable on site $i$, $\langle i,j\rangle$ denotes the nearest neighbors,
and $J>0$ is the coupling constant. In the notation of Gefen {\em et al.}\cite{gab84}, the 
Sierpi\' nski carpet studied here was generated by starting from the first order carpet consisting of $b\times b$ 
sites, with $b=$ 3, from which $l^{2}$ centrally located sites, $l=$ 1, were removed. Then the carpet of order $m$  
was obtained from the carpet of order $m-$1 by translating it with seven translation vectors as illustrated in 
Figure~\ref{fig1}. Since the outer edge of the $m$th order carpet is identical to the outer edge of a  
3$^{m}\times$3$^{m}$ square lattice, it is immediately clear that $R=\infty$ for the Sierpi\' nski carpets.
The fractal dimension of our carpet is $D=\ln(b^{2}-l^{2})/\ln b=\ln$8$/\ln$3 = 1.8928, and its connectivity $Q$, 
which is defined as the smallest of the fractal dimensions of the boundaries of the bounded subsets of the 
carpet, is $Q=\ln(b-l)/\ln b=\ln$2$/\ln$3 = 0.6309, see Ref.~\refcite{gab84}. The defining property of all fractals is 
that they are scale invariant but not translationally invariant. The lacunarity $L$ of a fractal measures its  
deviation from translational symmetry and the amount of inhomogeneity. Using an approximation 
for $L$ proposed by Gefen {\em et al.}\cite{gab84} we find $L=$ 0.0988\cite{lacunar}. This is much smaller  
than the values for two $b=$ 7 and $l=$ 3 Sierpi\' nski carpets considered in Ref.~\refcite{gab84} 
(for translationally invariant systems $L=$ 0\cite{frr94}).     

We computed the heat capacity, the susceptibility and the helicity modulus for Sierpi\' nski carpets of orders 
$m=$ 2--5 (the number of sites in the carpet of order $m$ is $N=$ 8$^{m}$, and we take the smallest $m=$ 1 carpet  
to have lattice spacing equal to one). The dependence of the helicity modulus on the size of the carpet, and on 
the boundary conditions, clearly indicates that in the thermodynamic limit there is no BKT transition      
at any finite temperature.

The rest of the paper is organized as follows. In Section~\ref{numerics} we outline the numerical procedure used
in calculations and then we present and discuss our numerical results in Section~\ref{results}. Finally, 
in Section~\ref{summary} we give a concise overview of our findings.

\section{Calculation}\label{numerics}

We have used Monte Carlo simulations based on the Wolff cluster algorithm\cite{wolff}, 
which avoids the critical slowing down associated with a diverging correlation length in the vicinity of a 
phase transition. One MC step involved generating a cluster of correlated spins as described in Ref.~\refcite{wolff}, 
and then flipping those spins about the randomly chosen reflection axis for that step. For each temperature
we discarded the first 120,000 MC steps, which allowed for the system to equilibrate. We then generated an  
additional seven links, each of 120,000 MC steps, and the final configuration after computing these 
seven links was used as the initial configuration for the next higher temperature. We estimated the error
in our calculations by breaking up each of the seven links into six blocks of 20,000 MC steps, then
calculating the average values for each of the 42 blocks and finally taking the standard deviation $\sigma$ of 
these 42 average values as an estimate of the error.

Since the characteristic feature of fractals is scale invariance (and not translational invariance),  
we could not employ the periodic boundary conditions. Instead, we used just two types of boundary
conditions: open (or free) and closed. For the closed boundary condition, each of the four outer corners  
of an $m$th order carpet is coupled to the closest two of the remaining three outer corners, while 
in the open boundary condition none of the outer corners are coupled to each other.  

The heat capacity per site $C$, and the linear susceptibility per site $\chi$, were computed in 
a standard way using the fluctuation theorems, as in our previous Monte Carlo studies\cite{mb10,pm11}.
From the fluctuations in energy, $C$ was calculated as follows:
\begin{equation}
	\label{eq:C}
	C=\frac{1}{N} \frac{\left<H^2\right>-\left<H\right>^2}{k_{B}T^2}\>,
\end{equation}
and $\chi$ was computed in a similar manner, using the fluctuations in magnetization per site ($m$):
\begin{equation}
	\label{eq:chi}
	\chi=\frac{\left<m^2\right>-\left<m\right>^2}{k_{B}T}.
\end{equation}
In equations~(\ref{eq:C}) and~(\ref{eq:chi}) $k_{B}$ is the Boltzmann constant, $T$ is the absolute 
temperature, and $\left < \cdots \right >$ denotes the MC average. To calculate the helicity modulus $\gamma$
we used the same procedure as in our earlier work, which was first proposed by Ebner and Stroud\cite{es83}. 
In this procedure, the XY Hamiltonian (equation~(\ref{eq:ham})) is thought of as describing a set of
Josephson-coupled superconducting grains in zero magnetic field. The variable $\theta_{i}$, which gives the
direction of spin on the lattice site $i$, becomes the phase of the superconducting order parameter for that site. 
An applied uniform vector potential $\bm{A}$ causes a shift in the phase difference $\theta_{i}-\theta_{j}$  
of the XY Hamiltonian by the amount 2$\pi\bm{A}\cdot(\bm{r}_{j}-\bm{r}_{i})$$/\Phi_{0}$, with $\bm{r}_{i}$ 
the position vector of site $i$, and $\Phi_{0}=hc/$2$e$ the flux quantum. The helicity modulus $\gamma$ is given   
by the second derivative of the Helmholtz free energy per site with respect to uniform $\bm{A}$, at $\bm{A}=$ 0. One finds  
\begin{equation}
 \gamma=\frac{1}{N}\left[\left<\left(\frac{\partial ^{2}H}{\partial A^{2}}\right)_{A=0}\right> 
        -\frac{1}{k_{B}T}\left<\left(\frac{\partial H}{\partial A}\right)^{2}_{A=0}\right> +
        \frac{1}{k_{B}T}\left<\left(\frac{\partial H}{\partial A}\right)_{A=0}\right>^{2}\right]\>, 
\end{equation}
\noindent with H the phase-shifted XY Hamiltonian. By applying a uniform vector potential along  
one of the edges of the Sierpi\' nski carpet (e.g. $\bm{A}$ along the $x$-axis), only nearest neighbors whose
$x$-coordinates differ by $\pm$1 (the smallest lattice spacing) will contribute to $\gamma$, as a 
consequence of the square symmetry of the carpet.

\section{Results and Discussion}\label{results}

\begin{figure}[ph] 
\centerline{\psfig{file=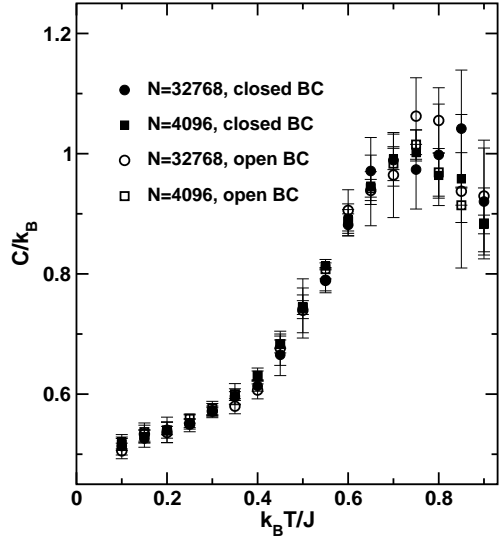,width=8cm}}
\vspace*{8pt}
\caption{Calculated heat capacity per site as a function of temperature and system size with both types of boundary
conditions.}
\label{fig2}
\end{figure}

Our results for the heat capacity per site for both the open and closed boundary conditions are 
presented in Figure~\ref{fig2}. 
The increasing size of the error bars with increasing temperature is a consequence of the fact that
the size of the clusters of correlated spins in the Wolff algorithm decreases with increasing temperature.
Hence, a larger number of Monte Carlo steps is required at higher temperatures. The results shown in
Figures~\ref{fig2}, \ref{fig4}, and \ref{fig6} at $k_{B}T/J\geq$ 0.55 were obtained by using 600,000 MC steps per link 
instead of 120,000 MC steps per link, which reduced the size of the error bars by about a factor of two.
%In simulations on systems that exhibit the continuous phase transition, such as the $XY$-model on a cubic
%lattice, the maximum in specific heat increases with the system size (see, for example, Figure 2 in 
%Ref.~\refcite{pm11}) 
%as a result of diverging correlation length at $T_{c}$ in an infinite system. Thus our results for $C$ in
%Figure \ref{fig2} indicate no long-range order at any finite temperature. This is in agreement with the
%conjecture by Gefen {\em et al.}\cite{gab84} based on the correspondence between pure resistor network and
%$n$-component spin models with $n\geq$2 for the Sierpi\' nski carpet with the fractal dimension $D$=1.8928.
%
%On the other hand, the classical $XY$-model on two-dimensional regular lattices displays BKT transition  
\newpage

We have examined the dependence of the maximum in the specific heat $C^{\mbox{max}}$ on the system size, Figure 3. 
The data on $C^{\mbox{max}}$ obtained with open boundary condition were fitted with the formula 
\begin{equation}
\label{eq:cvmax}
C^{\mbox{max}}(N)=C^{\infty}+\frac{Q}{(\ln N)^{a}}\>,
\end{equation}
where N is the number of sites in the carpet. We obtained $C^{\infty}=$ 1.134, $Q=$ -0.858 and $a=$ 0.86 with the 
$\chi^{2}$ of the fit equal to 5.938$\times$10$^{-6}$.
The maximum in the specific heat $C$ saturates with increasing system size.
The classical $XY$-model on two-dimensional regular lattices displays BKT transition 
from quasi-long-range order (order parameter correlation function decays algebraically) to disordered phase 
(order parameter correlation function decays exponentially). The transition results from unbinding of topological 
defects (vortices and antivortices) and the specific heat has an unobservable essential singularity at the transition 
temperature $T_{c}$\cite{bn79}. The maximum in heat capacity is above the transition temperature and is caused 
by unbinding of vortex clusters with increasing temperature above $T_{c}$\cite{tc79}. As a result it
saturates with increasing system size in Monte Carlo simulations\cite{tc79,vhc81}. 
While the saturation in $C$ with increasing system
size is not sufficient to prove the existence of BKT transition at finite temperature, it is a necessary
consequence if this transition does take place in an infinite system.\\

\vspace{0.2in}

\begin{figure}[ph]
\centerline{\psfig{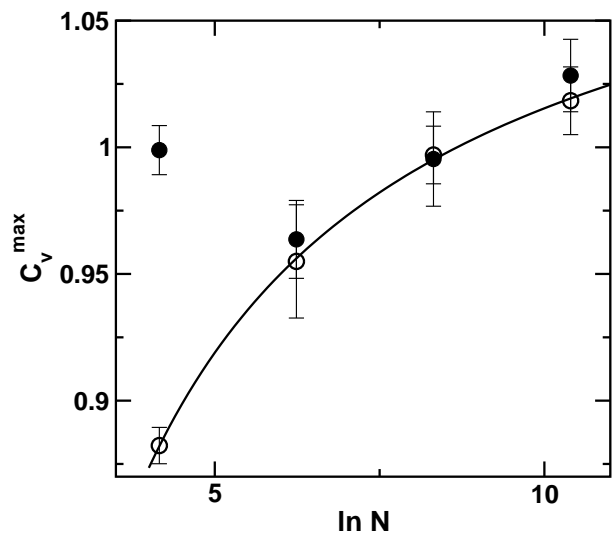}}
\vspace*{8pt}
\caption{The maximum in the specific heat as a function of the system size for closed boundary condition (filled
circles) and open boundary condition (open circles). The solid line is a fit to the results with open boundary
condition obtained with equation (\ref{eq:cvmax}).}
\label{fig3}
\end{figure}

We would like to point out that while in MC simulations of the classical $XY$-model on the square 
lattice\cite{vhc81} (see also Figure 2 in Ref.~\refcite{mb10}) the temperature $T^{*}$ at which the specific 
heat has maximum $C^{\mbox{max}}$ slowly decreases with increasing system size, we find the opposite trend for 
the Sierpi\' nski carpet. For the carpets with $N=$ 64, 512, 4096, and 32768 we obtained $T^{*}/J=$ 0.6, 0.7, 0.75, and 
0.75, respectively, with open boundary condition, and $T^{*}/J=$ 0.7, 0.7, 0.7, and 0.75, respectively, with closed 
boundary condition. 

Our results for the susceptibility are given in Figure~\ref{fig4}. 
The trend in the change of shape in $\chi$ with increasing system size is  
similar to what we obtained for the classical $XY$-model on the square lattice\cite{mb10}: the peaks
in $\chi$ increase in height and move to lower temperatures with increasing system size.  
We saw this behavior in both the Sierpi\' nski gasket\cite{mb10} and Sierpi\' nski pyramid\cite{pm11}, 
but with a more substantial shift in the peak position with increasing system size. For these finitely
ramified fractals, we concluded that in the thermodynamic limit, there was no finite-temperature transition.  
For the BKT transition, Kosterlitz predicted\cite{k74} that above the  transition temperature $T_{c}$ the
susceptibility diverges as $\chi\sim\exp[(2-\eta)b(T/T_{c}-1)^{-\nu}]$, with $\eta=$ 0.25, $b\approx$ 1.5,   
and $\nu=$ 0.5, and is infinite below $T_{c}$. For finite systems one gets finite peaks in $\chi$ above
$T_{c}$. 
%In view of our results for the heat capacity, as well as the trends in helicity modulus explained
%below, the peak in the susceptibility must continue to migrate toward zero temperature with increasing  
%system size, so that in the thermodynamic limit, there can be no finite-temperature divergence in $\chi$.
\begin{figure}[ph]
\centerline{\psfig{file=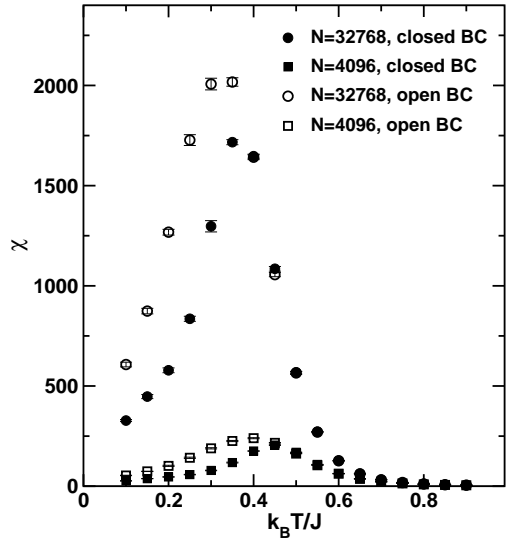,width=8cm}}
\vspace*{8pt}
\caption{Calculated linear susceptibility as a function of temperature and system size with both types of boundary
conditions.}
\label{fig4}
\end{figure}
In Figure~\ref{fig5} we show how the temperature $T_{\chi}$ corresponding to the maximum in the susceptibility 
varies with the system size for the case of closed boundary condition. We find that $T_{\chi}$ decreases linearly with 
$\ln N$. The fit to the data according to the power law
\begin{equation}
\label{eq:fitChi1}
T_{\chi}(N)=R+S(\ln N)^{b}
\end{equation}
gave $R=$ 0.85, $S=$ -0.048, and $b=$ 1 with the $\chi^{2}$ of the fit equal to 1.6$\times$10$^{-9}$ (solid line in
Figure~\ref{fig5}). This would imply that $T_{\chi}$ would become 0 for $N=$ 4.7$\times$10$^{7}$, which 
practically corresponds to the thermodynamic limit. We also fitted the data for $T_{\chi}$ according to 
\begin{equation}
\label{eq:fitChi2}
T_{\chi}(N)=U+\frac{W}{N^{c}}
\end{equation}
(i.e.~an exponential fit in $\ln N$) and obtained $U=$ -0.238, $W=$ 1.17, and $c=$ 0.0646 with the $\chi^{2}$ of the 
fit equal to 1.9$\times$10$^{-4}$ (dashed line in Figure~\ref{fig5}). The negative value of $U$ implies that 
$T_{\chi}$ is not positive in the thermodynamic limit, i.e.~that there is no finite-temperature BKT transition 
in the thermodynamic limit (equation (\ref{eq:fitChi2}) gives $T_{\chi}=$ 0 for $N=$ 5.1$\times$10$^{10}$, which 
again corresponds to the thermodynamic limit).  \\ 

\vspace{0.2in}

\begin{figure}[ph]
\centerline{\psfig{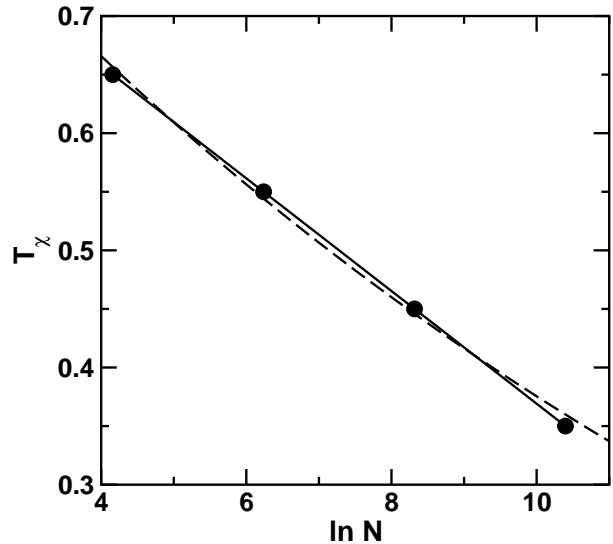}}
\vspace*{8pt}
\caption{The dependence of the temperature where the susceptibility attains the maximum on the system size. The 
solid line gives a fit to the data according to equation (\ref{eq:fitChi1}) and the dashed line gives a fit 
according to equation (\ref{eq:fitChi2}).}
\label{fig5}
\end{figure}

The best indicator of the BKT transition in the numerical work is the temperature dependence of the
helicity modulus $\gamma$ which measures the stiffness of the angles $\{\theta_{i}\}$ with respect to a twist 
at the boundary of the system. At zero temperature, when the angles are all aligned, the value of $\gamma$  
is finite, and at sufficiently high temperature, when the system is in the disordered paramagnetic phase, $\gamma=$ 0. 
In the case of the classical XY-model on three-dimensional regular lattices $\gamma(T)$ decreases  
continuously with increasing temperature and just below the transition temperature $T_{c}$ it obeys  
a power law $\gamma(T)\propto |T-T_{c}|^{v}$ (Ref.~\refcite{lt89}). In two dimensions Nelson and  
Kosterlitz\cite{nk77} predicted a discontinuous jump in $\gamma$ at the BKT transition temperature $T_{c}$  
with a universal value $\gamma(T_{c})/T_{c}=$ 2/$\pi$. For a finite system the jump in $\gamma$ is   
replaced by continuous decrease with increasing $T$, which becomes steeper near $T_{c}$ as the system  
size increases. 

Our results for $\gamma$ are shown in Figure~\ref{fig6} for the open and closed boundary conditions.  
\noindent 
\noindent The two important similarities between these results and the 
ones obtained for fractals with finite order of ramification\cite{mb10,pm11} are: (1) In both classes of fractals, 
the open boundary condition led to $\gamma=$ 0, within the error bars, at finite 
\begin{figure}[ph]
\centerline{\psfig{file=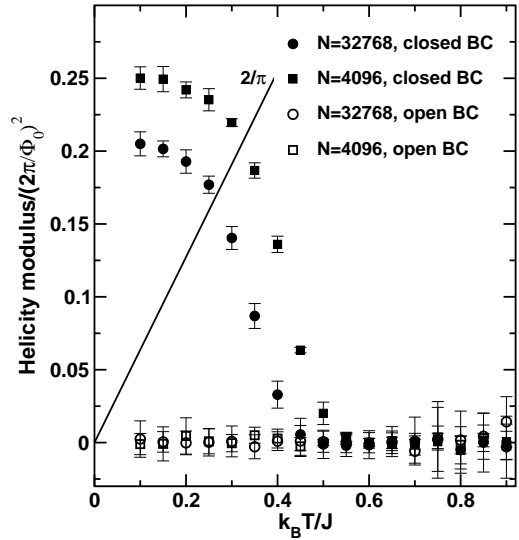,width=8cm}}
\vspace*{8pt}
\caption{Calculated helicity modulus as a function of temperature and system size with both types of boundary
conditions.}
\label{fig6}
\end{figure}
\noindent temperatures. (2) In each case, the closed boundary condition led to finite low-temperature values of $\gamma$ 
which decrease with increasing system size. Moreover, the onset of the downturn in $\gamma$, which is in 
the vicinity of the universal 2/$\pi$-line, moves to the lower temperatures with increasing system size. 
In the case of the $XY$-model on regular lattices (see Figure 4 in\ Ref.~\refcite{mb10} for the results 
obtained on the square lattices), low-temperature values of the helicity modulus do not depend on the system size.
Furthermore, the onset of the downturn in $\gamma$ is not size dependent, and it occurs at the same temperature 
for each square lattice considered.  

\begin{figure}[ph]
\centerline{\psfig{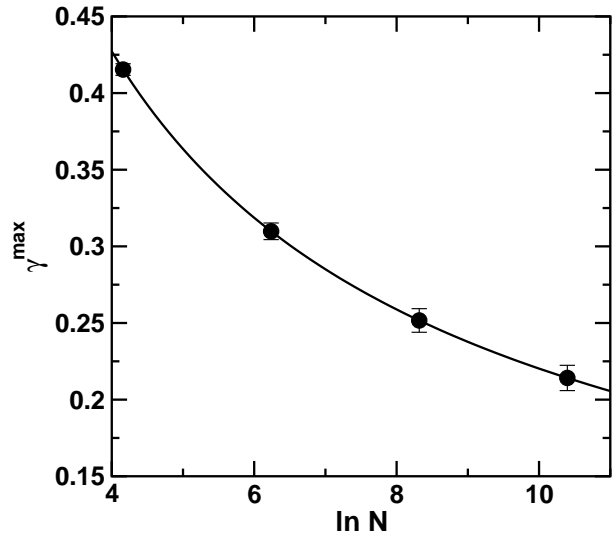}}
\vspace*{8pt}
\caption{The low-temperature value of the helicity modulus calculated with closed boundary condition as a 
function of the system size.}
\label{fig7}
\end{figure}

In Figure~\ref{fig7} we show how the low-temperature value of the
helicity modulus obtained with closed boundary condition depends on the system size.   
%As a result, we conclude that Monte Carlo simulation results for the
%classical $XY$-model on the  Sierpi\' nski carpet with $b$=3 and $l$=1 indicate the absence of finite
%temperature BKT transition in the thermodynamic limit.  
The data were fitted according to 
\newpage
\begin{equation}
\label{eq:fitHel}
\gamma^{\mbox{max}}(N)=X+\frac{Y}{(\ln N)^{f}}
\end{equation}
with $X=$ 2.67$\times$10$^{-9}$, $Y=$ 1.16, and $f=$ 0.723, and the $\chi^{2}$ of the fit was 8.8$\times$10$^{-20}$. 
A fit  where $X$ was set equal to 0 gave the same values for $Y$ and $f$ with the $\chi^{2}$ of 

\vspace{0.3in}

\noindent the fit equal to 4$\times$10$^{-4}$. Clearly the low-temperature value of the helicity modulus obtained 
with closed boundary condition goes to zero  
in the thermodynamic limit. As a result, we conclude that the Monte Carlo simulation results for the
classical $XY$-model on the  Sierpi\' nski carpet with $b=$ 3 and $l=$ 1 indicate the absence of finite 
temperature BKT transition in the thermodynamic limit.

\section{Conclusions}\label{summary}

We have performed an extensive Monte Carlo study of the classical $XY$-model on the Sierpi\' nski  
carpet with $b=$ 3 and $l=$ 1\cite{gab84}, which has infinite order of ramification $R$, fractal
dimension $D=$ 1.8928, connectivity $Q=$ 0.6309, and lacunarity $L=$ 0.0988. Our results for the helicity 
modulus obtained with closed boundary conditions for a given order carpet exhibit temperature dependence
similar to what we expect in square lattices undergoing BKT transition: a continuous drop near the 2$/\pi$ line. 
However, the dependence of the low-temperature values of the helicity modulus on the system size, as well as 
on the boundary conditions, implies that there is no finite-temperature BKT transition in this system in the 
thermodynamic limit. The numerical results for the dependence of the temperature where the susceptibility attains 
the maximum value on the system size are completely 
consistent with this conclusion. The absence of finite-temperature BKT transition in this infinitely 
ramified planar fractal is likely related to the fact that its fractal dimension $D$ is less than 2. A 
further analysis along the lines of Ref. \refcite{vkb} for finitely ramified   
planar Sierpi\' nski gasket would be required to clarify this.
%These results agree with the assertions made by Gefen \emph{et al.}\cite{gab84}
%for continuous spin models on infinitely ramified Sierpi\' nski carpets with fractal dimension $D<$2.

\section*{Acknowledgments}
This work was supported in part by the Natural Sciences and Engineering Research Council of Canada.
\\


\begin{thebibliography}{20}
\bibitem[11]{lacunar}%11
For $b=$ 3 and $l=$ 1, the points of the first order carpet are on a 3$\times$3 grid. When the carpet is viewed  
from any of the 8 sites which are on the boundary of the 3$\times$3 grid one sees 7 non-eliminated points
of the carpet, and the view from the 9th (central) site reveals 8 non-eliminated points. Hence the average 
number of non-eliminated points is $\bar{n}=$ (8$\cdot$7+8)/9 = 64/9, and their mean square deviation is
$L=$ [8(7-64/9)$^{2}$+(8-64/9)$^{2}$]/9 = 8/81.
\end{thebibliography}
\end{document}